\documentclass{svjour3}                     % onecolumn (standard format)
\smartqed  % flush right qed marks, e.g. at end of proof
\usepackage{graphicx}
\newcommand*{\LD}[2][N]{#1L$_#2$}
\newcommand*{\Mpp}{M_{\pi\pi}}

\journalname{Few-Body Systems (FB20)}
\begin{document}

\title{Gauge-invariant theory of two-pion photo- and electro- production off the
nucleon%
\thanks{Presented at the 20th International IUPAP Conference on Few-Body
        Problems in Physics, 20 - 25 August, 2012, Fukuoka, Japan}
}


\author{H. Haberzettl \and K. Nakayama \and Yongseok Oh}


\institute{Helmut Haberzettl \at
              %Institute for Nuclear Studies and
              Department of Physics, The George Washington University, Washington, DC 20052, USA \\
              \email{helmut.haberzettl@gwu.edu}           %  \\
        \and
    Kanzo Nakayama \at
    Department of Physics and Astronomy, University of Georgia,
    Athens, GA 30602, USA \\
    \email{nakayama@uga.edu}
        \and
    Yongseok Oh \at
    Department of Physics, Kyungpook National University,
    Daegu 702-701, Korea \\
    \email{yohphy@knu.ac.kr}
}

\date{Received: date / Accepted: date}

\maketitle

\begin{abstract}
A field-theoretical description of the photoproduction of two pions off the
nucleon is presented that applies to real as well as virtual photons in the
one-photon approximation. The Lorentz-covariant theory is complete at the level
of all explicit Faddeev-type three-body final-state mechanisms of dressed
interacting hadrons, including those of the nonlinear Dyson-Schwinger type. All
electromagnetic currents are constructed to satisfy their respective
(generalized) Ward-Takahashi identities and thus satisfy local gauge invariance
as a matter of course. The Faddeev-type ordering structure results in a natural
expansion of the full two-pion photoproduction current $\Mpp^\mu$ in terms of
multiple loops that preserve gauge invariance order by order in the number of
loops, which in turn lends itself naturally to practical applications of
increasing sophistication with increasing number of loops.
%Insert your abstract here. Include keywords, PACS and mathematical
%subject classification numbers as needed.
\end{abstract}

\section{Introduction}
\label{intro}

The experimental study of double-pion production off the nucleon has a fairly
long history, with some of the earliest experiments going back to more than
half a century. In the last two decades, with the availability of sophisticated
experimental facilities at MAMI in Mainz, GRAAL in Grenoble, ELSA in Bonn, and
the CLAS detector at JLab, the emphasis of experiments with both real and
virtual photons is clearly on using this reaction as a tool to study and
extract the properties of excited hadronic states that form at intermediate
stages of the reaction.

Theoretically, the study of double-pion photo- and electroproduction off the
nucleon is a challenging problem because, unlike single-pion production, its
correct description necessarily must combine baryonic and mesonic degrees of
freedom on an equal footing since the two final pions can come off a decaying
intermediate meson state, and not just off intermediate baryons as a sequence
of two single-pion productions. This requires accounting for all competing
internal photo-subprocesses like, for example, the baryonic $\gamma N\to \pi N$
and the purely mesonic $\gamma\rho\to\pi\pi$ in a consistent manner. Moreover,
this means that for double-pion production the entire field of meson
spectroscopy becomes an integral part of the problem, in addition to all
baryon-spectroscopic issues well known from single-pion production.

%
%=======================================================
\begin{figure*}[t!]\centering
\includegraphics[width=.65\textwidth,clip=]{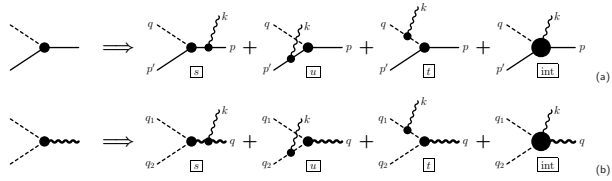}
  \caption{\label{fig:Msutc}%
  Generic topological structures of the single-pion production current for
  $\gamma N\to \pi N$ and of two-pion production off a single meson, here taken
  to be the $\rho$, i.e., $\gamma \rho\to \pi\pi$, resulting from attaching a
  photon to the respective hadronic vertex on the left. The labels $s$, $u$,
  $t$ refer to the Mandelstam variables of the respective intermediate
  off-shell hadron, and ``int" labels the interaction currents resulting from
  the photon interacting with the interior of the respective vertex (which also
  subsume the final-state interactions of the respective outgoing hadronic
  two-body systems~\cite{Haberzettl1997,HNK2006}). Time proceeds from right to
  left in all diagrams. }
\end{figure*}
%=======================================================
%

\section{Formalism}

We follow here the general field-theoretical approach of Haberzettl outlined in
Ref.~\cite{Haberzettl1997} for single-pion photoproduction off the nucleon,
which is based on an LSZ-type reduction scheme where the photon is coupled to
the connected part of the underlying hadronic Green's function by the
gauge-derivative formalism~\cite{Haberzettl1997}. The resulting generic
topological structures for four-point currents like $\gamma N\to \pi N$ and
$\gamma \rho\to \pi\pi$ are shown in Fig.~\ref{fig:Msutc}. The respective
interaction currents subsume the full complexity of the final-state
interactions of the outgoing two-hadron systems. Practical ways to deal with
this complexity are described in Ref.~\cite{HNK2006}. Using the reformulation
of the production current of Ref.~\cite{HHN2011}, this formalism was recently
employed successfully for calculating single-pion production
observables~\cite{Huang2011}.

\begin{figure}[t!]\centering
  \includegraphics[width=.4\columnwidth,clip=]{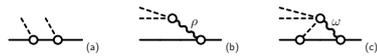}
  \caption{\label{fig:basicprocess}%
  Basic two-pion production processes: (a) sequential production along the
  nucleon line, and (b) intermediate production of a $\rho$-meson decaying into
  two pions.  Part (c) provides an example of loop mechanisms based on
  intermediate multi-meson vertices. In addition to $\rho$ and $\omega$, other
  intermediate mesons with two-pion and three-pion decay modes, respectively,
  are possible.  Possible contributions from five- or more-meson vertices are
  not shown.}
\end{figure}

\begin{figure*}[t!]\centering
  \includegraphics[width=.6\textwidth,clip=]{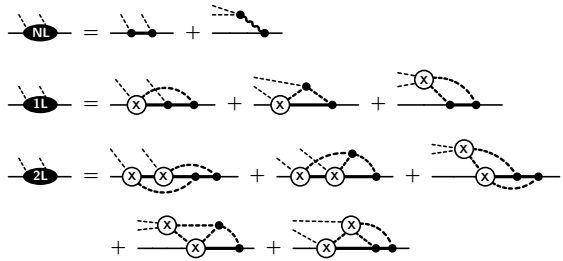}
  \caption{\label{fig:All2Pi}%
  Grouping of hadronic two-pion production mechanisms off the nucleon involving
  no loop (NL), one loop (1L), and two loops (2L). (All loops involve
  \emph{fully dressed} entities.) The thick interior lines subsume all
  particles permitted by the process, with the solid lines indicating baryons
  and the dashed lines mesons. The thick wavy line stands for those mesons
  (like $\rho$, $\omega$, etc.) that can decay into two pions (for intermediate
  mesons, such mesons are subsumed under the heavy dashed line). Summations over
  all permitted internal particles are implied. All vertices are fully dressed
  and the various meson-baryon or meson-meson scattering processes indicated by
  $X$ are \emph{non-polar}, i.e., they \emph{do not} contain $s$-channel
  driving terms because their contributions are already subsumed in the full
  dressing of the vertices (see also~\cite{Haberzettl1997}).}
\end{figure*}

The starting points for describing the reaction $\gamma N\to \pi\pi N$ are the
elementary hadronic two-pion production processes depicted in
Fig.~\ref{fig:basicprocess}. For the present note, for lack of space, we will
ignore here contributions arising from three- or more-pion vertices as depicted
in Fig.~\ref{fig:basicprocess}(c). Their detailed treatment will be explained
in a forthcoming publication~\cite{HNO2pi}. Employing the elementary
interactions to all orders then produces the fully dressed hadronic diagrams
shown in Fig.~\ref{fig:All2Pi}. These diagrams represent the lowest orders of a
three-body multiple scattering series that can be summed in closed form in
terms of the Faddeev-type ordering structure of the Alt-Grassberger-Sandhas
equations~\cite{AGS67}, amended by nonlinear driving terms that account for the
fact that at intermediate stages infinitely many mesons may be produced (for
full details, see~\cite{HNO2pi}).

%
%=======================================================
%=======================================================
%

%=======================================================
%=======================================================
%

%
%=======================================================
\begin{figure*}[t!]\centering
  \includegraphics[width=.5\textwidth,clip=]{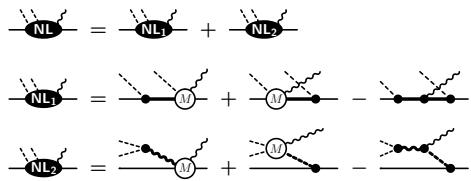}
  \caption{\label{fig:MNL}%
  Two-pion photoproduction at the no-loop level where the photon is attached to
  the NL diagrams of Fig.~\ref{fig:All2Pi}. The two contributions \LD{1} and
  \LD{2} correspond to the two NL diagrams in  Fig.~\ref{fig:All2Pi} in the
  order given. The photoproduction subamplitudes labeled $M$ each comprise
  four generic terms, similar to those shown in Fig.~\ref{fig:Msutc}. The
  subtractions correct the double counting resulting from the photon being
  attached to the respective intermediate particle in both preceding diagrams,
  i.e., when expanding all amplitudes $M$,
  each group consists of seven diagrams. The \LD{1} and \LD{2}
  diagram groups satisfy \emph{independent} gauge-invariance constraints.}
\end{figure*}
%=======================================================

%
%=======================================================
\begin{figure*}[t!]\centering
  \includegraphics[width=.61\textwidth,clip=]{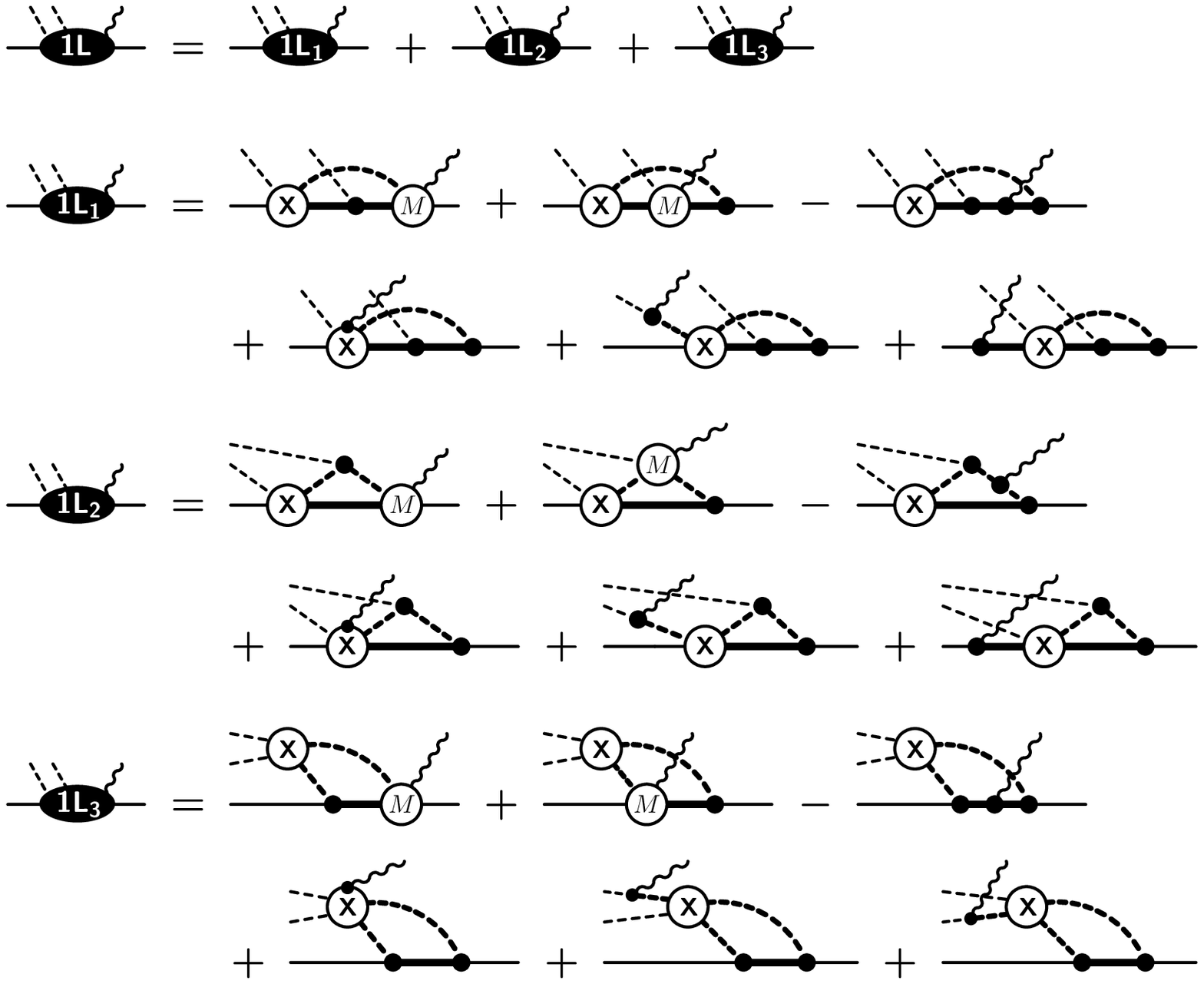}
  \caption{\label{fig:Loop1}%
  Two-pion-production currents resulting from coupling the photon to the 1L
  diagrams in Fig.~\ref{fig:All2Pi}. The subtractions correct double counting
  of the corresponding mechanisms. For an explanation of the five-point
  interaction currents $X^\mu$, see, for example, Fig. 6 in
  Ref.~\cite{Haberzettl1997}. Each group \LD[1]{i} ($i=1,2,3$) obeys an
  \emph{independent} gauge-invariance constraint. }
\end{figure*}
%=======================================================
%

Applying now the gauge-derivative formalism~\cite{Haberzettl1997} to the basic
\emph{dressed} hadronic processes of Fig.~\ref{fig:All2Pi} amounts to summing
up attaching the photon to these diagrams in all possible ways. For the no-loop
(NL) and one-loop (\LD{1}) diagrams, the resulting currents are shown in
Figs.~\ref{fig:MNL} and \ref{fig:Loop1}, respectively. Higher-loop orders are
not shown for lack of space. We emphasize that all input current subamplitudes
appearing in Figs.~\ref{fig:MNL} and \ref{fig:Loop1} maintain full \emph{local}
gauge invariance, as this is necessary for any reaction theory that aims to be
microscopically consistent; mere global gauge invariance is not sufficient in
this respect. The respective four-divergences for each of the input current
pieces, therefore, provide \emph{off-shell} generalized Ward-Takahashi
identities~\cite{Haberzettl1997}. It is shown in
Refs.~\cite{Haberzettl1997,HNK2006,HHN2011} how to do this even if some current
pieces cannot be calculated exactly (which in practice invariably will be the
case). Utilizing these gauge-invariant pieces as input in Figs.~\ref{fig:MNL}
and \ref{fig:Loop1}, one then shows easily that the resulting diagram groups
are \emph{separately} gauge invariant as a matter of course, with each group's
four-divergence providing a generalized Ward-Takahashi identity of its own.

\section{Summary}

Following the field-theoretic approach of Haberzettl~\cite{Haberzettl1997}, the
present note provides a diagrammatic expansion of the two-pion production
current off the nucleon in terms of currents resulting from topologically
distinct hadronic processes. Each current group satisfies full \emph{local}
gauge invariance provided the input currents individually satisfy their
respective generalized Ward-Takahashi
identities~\cite{Haberzettl1997,HNK2006,HHN2011}. The formalism thus provides a
microscopically consistent description of the reaction dynamics of two-pion
production off the nucleon. Full details will be given in a forthcoming
publication~\cite{HNO2pi}.

\begin{acknowledgements}
This work was supported in part by the National Research Foundation of Korea
funded by the Korean Government (Grant No.~NRF-2011-220-C00011). The work of
K.N. was also supported partly by the FFE Grant No.~41788390 (COSY-058).
\end{acknowledgements}

\end{document}